# Inverse Halftoning Through Structure-Aware Deep Convolutional Neural Networks

Chang-Hwan Son[0000-0001-7077-3074]

Kunsan National University, South Korea
cson@kunsan.ac.kr

**Abstract.** The primary issue in inverse halftoning is removing noisy dots on flat areas and restoring image structures (e.g., lines, patterns) on textured areas. Hence, a new structure-aware deep convolutional neural network that incorporates two subnetworks is proposed in this paper. One subnetwork is for image structure prediction while the other is for continuous-tone image reconstruction. First, to predict image structures, patch pairs comprising continuous-tone patches and the corresponding halftoned patches generated through digital halftoning are trained. Subsequently, gradient patches are generated by convolving gradient filters with the continuous-tone patches. The subnetwork for the image structure prediction is trained using the mini-batch gradient descent algorithm given the halftoned patches and gradient patches, which are fed into the input and loss layers of the subnetwork, respectively. Next, the predicted map including the image structures is stacked on the top of the input halftoned image through a fusion layer and fed into the image reconstruction subnetwork such that the entire network is trained adaptively to the image structures. The experimental results confirm that the proposed structure-aware network can remove noisy dot-patterns well on flat areas and restore details clearly on textured areas. Furthermore, it is demonstrated that the proposed method surpasses the conventional state-of-the-art methods based on the deep convolutional neural network, U-Net, and locally learned dictionaries.

## 1 Introduction

Digital halftoning is a process of generating a halftone image with homogenously distributed black and white dots from a continuous-tone image with discrete gray levels (e.g., 255 gray levels) [1]. It has been used primarily in bilevel output devices such as printers and copiers to render an image on a paper. In laser printers, black and white dots are used to control a laser beam to form a latent image on a photoconductor drum, and determines whether toner particles will touch the surface of the drum. Similarly, in inkjet printers, a halftoned image determines the spatial position of the ink that drops on a paper. Digital halftoning is used in other applications, for example, animated GIF generation from videos [2], removal of contour artifacts in displays [3], video processing in electronic papers [4], and data hiding [5]. The typically used digital halftoning are dithering, error diffusion, and direct binary search [6].

Inverse halftoning is the reverse of digital halftoning; in other words, a continuous-tone image with 255 gray levels or more is reconstructed from its halftoned version [7]. Inverse halftoning is required in several practical applications: bilevel data compression [8], watermarking [9], digital reconstruction of color comics [10], and high dynamic range imaging [11]. Digital halftoning is a many-to-one mapping; hence, inverse halftoning is an ill-posed problem with many possible solutions. Many approaches have been introduced over the last several decades based on look-up tables [12], adaptive low-pass filtering [13], maximum-a-posterior estimation [14], local polynomial approximation and intersection of confidence intervals (LPA-ICI) [15], and deconvolution [16]. Recently, machine-learning approaches have been actively studied based on dictionary learning [17-19] and neural networks [20]. In particular, deep convolutional neural networks (DCNNs) [21-24], which is a variant of neural networks, have demonstrated powerful performance for image restoration, segmentation, and classification problems. Hence, the DCNN can be used directly for inverse halftoning.

## 1.1 Primary Problem for Inverse Halftoning

Through digital halftoning, which includes binary quantization, image structures such as lines and regular patterns are lost inevitably on textured areas, whereas noisy dots are densely or sparsely generated on flat areas. Particularly, white dots that are distributed sparsely on dark areas, and black dots on bright areas resemble impulse-like noises. Therefore, the critical issue in inverse halftoning is removing noisy dots on flat areas and restoring image structures precisely from quantized binary data. Even though the DCNN is used for inverse halftoning, many problems are still unresolved. More specifically, the DCNN can automatically learn a large number of model parameters hierarchically from the given training patches. This implies that low-level features can be reused to represent high-level features through a layer-by-layer feature transformation; thus, the nonlinear mapping relationship between halftoned patches and continuous-tone patches can be predicted more accurately. The application of the DCNN to inverse halftoning can result in improvements in detail representation and dot elimination. However, further improvements are possible.

## 1.2 Proposed Approach

For image restoration [22], the layers typically used are the convolution layers with filters and the rectified linear unit (ReLU) layers, which are arranged in a sequence to build the architecture of the DCNN. The last layer is the loss layer that minimizes the mean square error (MSE) between the predicted continuous-tone patches and the original continuous-tone patches. Given the training data that comprise continuous-tone patches and halftoned patches, the mini-batch stochastic gradient descent algorithm in [25] is used to update the filters iteratively to finally complete the nonlinear transformation from the input layer to the last loss layer.

From the perspective of regression analysis, this training approach is similar to global regression, in which the parameters of a hypothesis function is learned without partitioning the training data into subsets. Meanwhile, local regression partitions training data into subsets with similar attributes, and subsequently learns the parameters of the hypothesis functions to be fitted into each subset. If this type of local regression approach is used for inverse halftoning [18,19], the detail representation can be improved.

Inspired by the method used in [18] that partitions training data into subsets, a new image structure map predictor (ISMP), which is the subnetwork to predict image structures, is introduced in this study. In addition, the method to combine the ISMP with the reconstruction subnetwork (RS) to recover continuous-tone images from the input halftoned images is presented. The primary idea is to predict the image structure map through the ISMP and connect the predicted map to the RS such that the entire network is trained in an end-to-end manner. This training strategy can provide useful information regarding which areas are smooth or textured. Hence, the entire network can be trained adaptively to local image structures. The predicted image structure map serves as a guide for a more accurate image reconstruction.

### 1.3 Contributions

- In this study, a new structure-aware DCNN is proposed for inverse halftoning. Specifically, the method to design three subnetworks and subsequently combining them to be trained in an end-to-end manner is described. The first subnetwork is the initial reconstruction subnetwork (IRS), which aims to generate initial continuous-tone image from an input halftoned image. The second subnetwork is the ISMP by which the initial continuous-tone image is transformed into an image structure map. The third subnetwork is the RS to reconstruct the final continuous-tone image from three types of images: halftoned image, image structure map, and initial continuous-tone image. The proposed architecture requires three types of images for continuous-tone image reconstruction, whereas the typically used DCNN architecture requires only one image (i.e., halftoned image). The three subnetworks are combined through a fusion layer to create the entire network and trained in an end-to-end manner. In the entire network, the image structure map output by the ISMP is fed into the RS to provide useful information about which areas are flat or textured. Hence, the entire network can be trained adaptively to image structures. Briefly, this study presents a method of predicting an image structure map from input halftoned images and demonstrates the method to train the entire network adaptively to local image structures. This is the primary contribution of this paper.
- If the ISMP is excluded, the proposed architecture becomes identical to the conventional DCNN. Through this study, it is verified whether the ISMP can increase the restoration quality through the performance comparison between the proposed structure-aware DCNN and the conventional DCNN. Furthermore, it is confirmed that the proposed method surpasses the conventional state-of-the-art methods: DCNN [22], U-Net [24], and LLD [18].

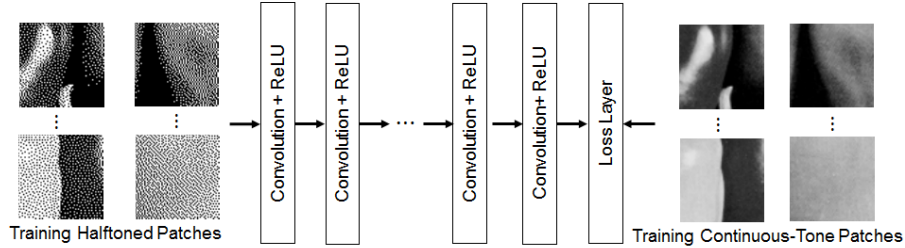

**Fig. 1.** Architecture of the deep convolutional neural network for inverse halftoning.

## 2   Conventional Approach based on DCNN

The typically used DCNN architecture for image restoration [22] can be applied to inverse halftoning, as shown in Fig. 1. The DCNN architecture consists of two types of layers: the convolution layers and ReLU layers, which are arranged in a sequence. The convolution layer convolves filters with input feature maps to extract local features, and the ReLU layer forces negative input values to be zero to consider the nonlinearity. The last layer is the loss layer to minimize the MSE between the predicted and original data. To train the DCNN in a supervised manner, halftoned patches and continuous-tone patches are required, and are fed into the input layer and loss layer, respectively, as shown in Fig. 1. The filters are initialized randomly, and subsequently updated iteratively using the mini-batch stochastic gradient descent algorithm to minimize loss. If a stop criterion is satisfied, the nonlinear transformation from the input layer to the last loss layer is completed. To test the DCNN, the last loss layer is removed and the halftoned image fed into the input layer is passed through the trained nonlinear transformation, thus finally producing a continuous-tone image. This DCNN-based approach has contributed significantly to the restoration quality. Nevertheless, further improvements are possible.

In Fig. 1, the layer-by-layer nonlinear transformation can be regarded as a black box that transforms the halftoned patches into continuous-tone patches. From the perspective of regression analysis, this training approach corresponds to global regression that learns the parameters of a hypothesis function without partitioning training data into subsets. It was reported in [18] that the local regression approach effectively improves the detail representation by fitting the hypothesis functions into subsets with similar image structures. In Fig. 1, the DCNN is trained without dividing the training data into subsets. Hence, the DCNN can lack detail representation and dot elimination.

## 3   Proposed Structure-Aware DCNN for Inverse Halftoning

Inspired by the method in [18], a new structure-aware DCNN architecture is proposed. The key idea is to provide the DCNN with useful information about local image structures. Fig. 2 shows the proposed structure-aware DCNN architecture for inverse halftoning. Three subnetworks exist in the proposed architecture. The first one is the IRS that aims to generate the initial continuous-tone image from an input halftoned image. The IRS comprises two types of layers: convolution and ReLU layers. It is noteworthy that no loss layer exists in the IRS. This implies that the IRS is pretrained. After training the IRS with the loss layer to minimize the MSE, the last loss layer is removed.

The second subnetwork is the ISMP that is connected to the back of the IRS to predict the image structure map. Digital halftoning includes binary quantization, and thus information loss inevitably occurs. Because halftoned images contain much less information than continuous-tone images, it is preferable to predict the image structures from the initially reconstructed continuous-tone image than from the halftoned image. Hence, the feature map at the last layer of the IRS is fed into the input layer of the ISMP. The loss layer of the ISMP requires two inputs. One is the predicted gradient patch that corresponds to the feature map at the layer before the loss layer. The other is the original gradient patch. In this study, the Sobel gradient operator is applied to the original continuous-tone patch to generate the gradient patch. A few examples of gradient patches and the corresponding continuous-tone patches are provided at the right side of Fig. 2. It is confirmed that the gradient patches contain local image structures such as lines, curves, or textures.

The third subnetwork is the RS. In Fig. 2, the feature map at the layer before the loss layer of the ISMP is the predicted image structure map. The concatenation layer stacks three types of images to form a three-dimensional tensor that is fed into the input layer of the RS. The three types of inputs are the halftoned image, initial continuous-tone image, and image structure map. However, the proposed entire network in Fig. 2 is trained in an end-to-end manner, and thus the initial continuous-tone image cannot be preserved. In other words, even though the pretrained IRS starts by generating the initial continuous-tone image at the last layer, it adjusts the initial continuous-tone image through backpropagation such that the ISMP is more accurate. Therefore, the key idea of the proposed architecture is that the predicted image structure map is connected to the input layer of the RS. The information about the image structure map can teach the RS regarding which areas are flat, lined, or textured. This enables the entire network to be trained by adapting to local image structures.

If the IRS and ISMP are excluded from Fig. 2, the proposed architecture becomes identical to the conventional DCNN in Fig. 1. Therefore, the extent of restoration quality improvement with the ISMP can be verified by comparing the proposed structure-aware DCNN and the conventional DCNN. Even though several ways exist for combining the IRS and ISMP with the RS in Fig. 2, the primary focus herein is to demonstrate the effectiveness of the ISMP in improving the restoration quality. Hence, the ISMP is connected to the RS. In this study, the predicted image structure map is stacked on the top of the input halftoned image through the concatenation layer, and subsequently fed into the input layer of the RS. This architecture can clarify the primary difference between the proposed structure-aware DCNN and the conventional DCNN in Fig. 1.

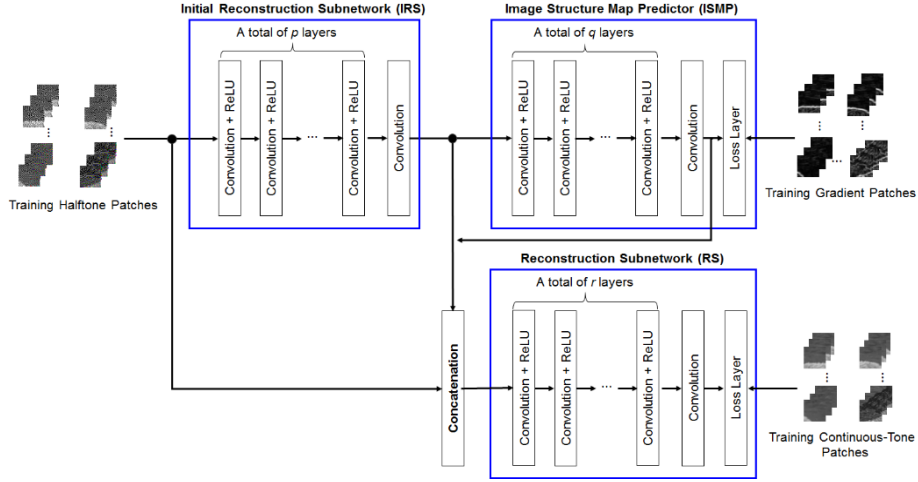

**Fig. 2.** Proposed structure-aware DCNN for inverse halftoning.

## 4 Experiments

The proposed structure-aware DCNN for inverse halftoning is implemented using MatConvNet [26] and trained with two 1080Ti GPUs on a Windows operating system. To compare the proposed method, state-of-the-art methods based on the LLD [18], DCNN [22], U-Net [24], and LPA-ICI [15] were tested. For performance evaluation, the peak signal-to-noise ratio (PSNR) and structure similarity (SSIM) [27] are used to measure the inverse of the MSE in a log space and the structure similarity between two images, respectively. In both of the PSNR and SSIM, a higher value indicates a higher quality. The training and testing codes of the proposed structure-aware DCNN method can be downloaded at https://sites.google.com/view/chson/home.

### 4.1 Training Data Collection

For training, public datasets [28] including General 100, Urban 100, BSDS100, and BSDS200 are used to prepare continuous-tone color images. The total number of continuous-tone color images is 500. For digital halftoning, the continuous-tone color images are converted into grayscale images, and subsequently error diffusion [29] is used to transform the grayscale images into halftoned images. The Floyd–Steinburg filter [1] is used for error diffusion. The Sobel gradient operator is applied to the grayscale images to obtain gradient images containing gradient magnitudes. To obtain the training patches, three types of patches are extracted randomly from the grayscale images, gradient images, and halftoned images. The patch size extracted is 32×32. In this study, grayscale patches are used for training because error diffusion can be easily applied to them.

**Table 1.** Number of filters and channels used in the convolutional layers.

| Subnetworks \ Layers | Input layer | Last convolutional layer | Other layers |
|---|---|---|---|
| IRS | $c=1$, $m=64$ | $c=64$, $m=1$ | $c=64$, $m=64$ |
| ISMP | $c=1$, $m=64$ | $c=64$, $m=1$ | $c=64$, $m=64$ |
| RS | $c=3$, $m=64$ | $c=64$, $m=1$ | $c=64$, $m=64$ |

### 4.2 Network Training

In the subnetworks, $m$ filters of size $5 \times 5 \times c$ are used in the convolution layers. Here, $c$ represents the number of input channels. Table 1 shows the number of filters and channels used in the convolution layers. In the input layer of the RS, $c$ is set to 3 because three input channels are fed into the input layer. The filters are initialized using a random number generator. The $p$, $q$, and $r$ numbers in Fig. 2 are set to 16, 6, and 16, respectively. To update the filters, the mini-batch gradient descent algorithm is used. The epoch number is 200 and the batch size is 64. Each epoch involves 1,000 iterations of backpropagation. The learning rate is $10^{-5}$. All the loss functions are modeled by the $l$-2 norm.

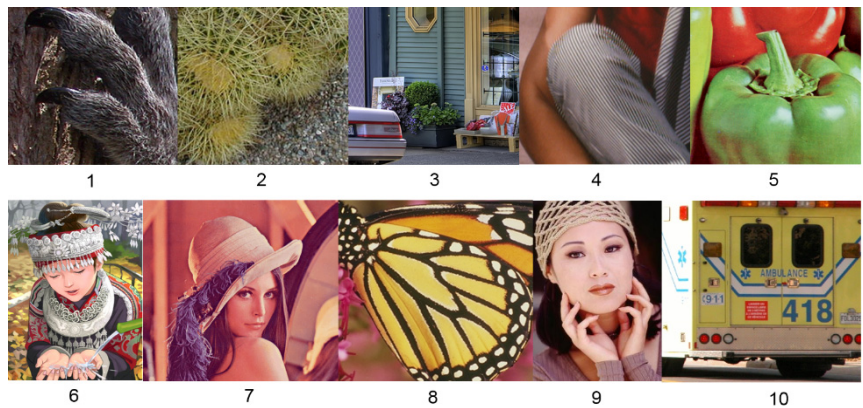

**Fig. 3.** Test images.

### 4.3 Performance Comparison

Fig. 3 shows the 10 test images for visual quality evaluation that are numbered accordingly. The test images contain various types of image structures including lines, curves, and regular patterns to verify whether the proposed structure-aware DCNN can improve detail representation and dot elimination. Fig. 4 shows the experimental results. As shown in the red boxes, the proposed method describes the image structures more accurately. In addition, the overall sharpness is better. Particularly, on the first row, more lines are restored with the proposed method. On the second row, the outlines of the grains of sand are restored more clearly and the textures on the palm and the hair accessory are described in more detail on the third row. On the fourth and last rows, the texts and the outline of the rip are more clearly restored, respectively. Moreover, the proposed method removes noisy dots on the flat areas completely, as shown in the blue box of the last row, which is not the case with the conventional DCNN [22] and U-Net [24] methods. By comparing the proposed method and the DCNN-based method [22], it is verified that the additional use of the ISMP can increase the performance for detail representation and dot elimination. From these results, it is concluded that the ISMP subnetwork enables the entire network to be trained adaptively to local image structures.

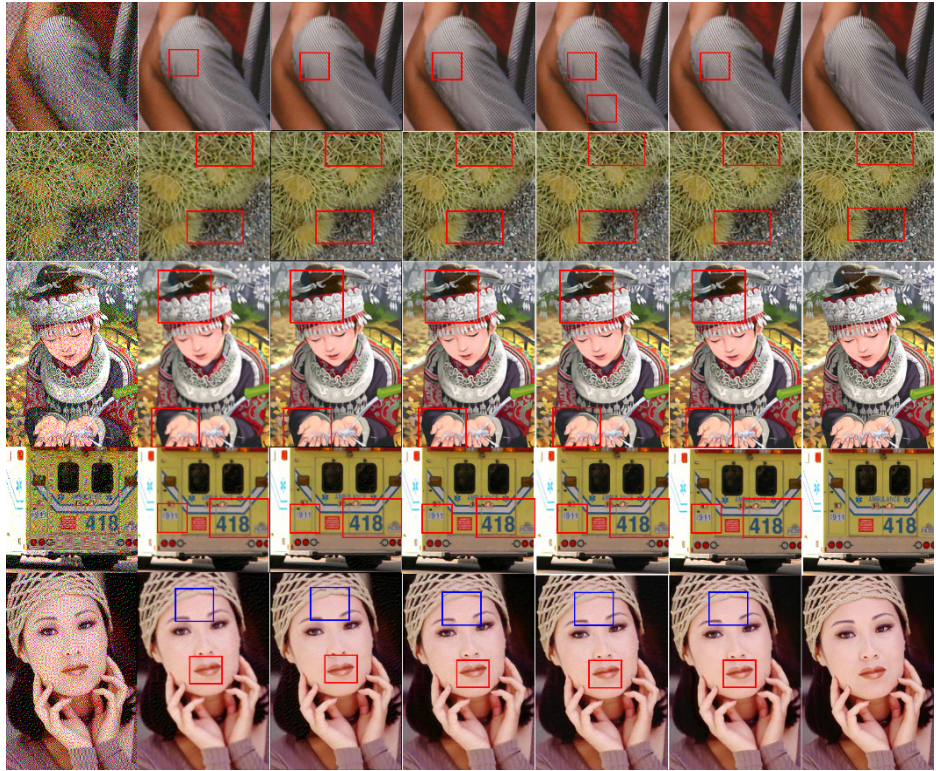

**Fig. 4.** Experimental results: halftoned images (first column), reconstructed images with LPA-ICI [15] (second column), reconstructed images with LLD [18] (third column), reconstructed images with DCNN [22] (fourth column), reconstructed images with U-Net [24] (fifth column), reconstructed images with proposed method (sixth column), and original images (last column).

Table 2 shows the results of the PSNR and SSIM evaluations. As expected, the proposed method demonstrates the best performance among all the methods and particularly surpasses the conventional DCNN and U-Net methods. This implies that the ISMP provides the RS with information regarding the areas that are smooth or textured, thus enabling the entire network to be trained adaptively to local image structures. In other words, the ISMP serves as a guide for a more accurate image reconstruction. The performance of the LLD method does not surpass that of the DCNN-based method. Before the emergence of the DCNN, the LLD method demonstrated one of the best performances. However, this method represents input patches using the linear combination of signal-atoms and sparse coefficients. Hence, modeling the nonlinearity is restricted, i.e., binary quantization occurs inevitably during digital halftoning. Meanwhile, the DCNN learns a large number of parameters hierarchically. Low-level features can be reused to represent high-level features through a layer-by-layer feature transformation. Even a nonlinear system can be modeled more accurately. The LPA-ICI belongs to nonlinear filtering, which fuses directional estimates. However, the model capacity is insufficient to preserve the overall sharpness. As shown in Table 2, the averaged PSNR value of the LPA-ICI method is the lowest among all the methods. In Table 2, it is shown that the performance of the U-Net-based method [24] is better than that of the DCNN-based method [22]. The U-Net downsamples the input image through the encoder and then upsamples the feature maps with skip connections through the decoder, which means that this network includes the multiscale fusion like as wavelets, which can lead to more improvement in the restoration quality.

**Table 2.** Performance evaluation.

| Methods | Proposed Method | | U-Net [24] | | DCNN [22] | | LLD [18] | | LPA-ICI [15] | |
|---|---|---|---|---|---|---|---|---|---|---|
| Test Images | PSNR | SSIM | PSNR | SSIM | PSNR | SSIM | PSNR | SSIM | PSNR | SSIM |
| 1 | 25.659 | 0.824 | 25.563 | 0.815 | 25.181 | 0.808 | 25.016 | 0.799 | 23.975 | 0.749 |
| 2 | 25.633 | 0.907 | 25.590 | 0.904 | 25.395 | 0.900 | 25.541 | 0.818 | 24.031 | 0.857 |
| 3 | 25.473 | 0.866 | 25.247 | 0.857 | 24.81 | 0.846 | 24.654 | 0.798 | 23.485 | 0.794 |
| 4 | 29.410 | 0.882 | 29.262 | 0.873 | 28.608 | 0.854 | 28.500 | 0.846 | 27.034 | 0.795 |
| 5 | 32.291 | 0.982 | 31.901 | 0.979 | 31.818 | 0.979 | 31.023 | 0.87 | 31.163 | 0.979 |
| 6 | 25.916 | 0.905 | 25.820 | 0.899 | 25.370 | 0.890 | 25.15 | 0.869 | 24.233 | 0.858 |
| 7 | 31.578 | 0.981 | 31.248 | 0.979 | 31.084 | 0.979 | 30.514 | 0.844 | 30.457 | 0.976 |
| 8 | 28.273 | 0.968 | 27.992 | 0.966 | 27.275 | 0.959 | 27.563 | 0.919 | 25.889 | 0.950 |
| 9 | 30.886 | 0.962 | 30.539 | 0.948 | 30.237 | 0.949 | 30.013 | 0.874 | 28.924 | 0.920 |
| 10 | 29.901 | 0.937 | 29.601 | 0.930 | 29.214 | 0.928 | 28.645 | 0.865 | 27.615 | 0.900 |
| AVG. | **28.502** | **0.921** | 28.276 | 0.915 | 27.899 | 0.909 | 27.662 | 0.850 | 26.681 | 0.878 |

## 5   Conclusion

A new structure-aware DCNN for inverse halftoning was proposed. The central idea was to build the subnetwork for image structure prediction and combine it with another subnetwork to be trained in an end-to-end manner for continuous-tone image reconstruction. The experimental results confirmed that the proposed architecture enabled the entire network to be trained adaptively to local image structures, and thus noisy dot-patterns on the flat areas were removed completely and local image structures such as lines and patterns were described precisely. It was also demonstrated that the proposed method yielded a better performance than the state-of-the art methods: DCNN, U-Net, and LLD.

## Acknowledgment

This work was supported by the National Research Foundation of Korea (2017 R1D1A3B03030853).